\documentclass[aps,prb,twocolumn,superscriptaddress,showpacs]{revtex4-1}
\usepackage{graphicx}
\pdfoutput=1

\newcommand{\mub}{$\mu_B$~}
\newcommand{\caxis}{$\textit{\textbf{c}}$~}
\newcommand{\aaxis}{$\textit{\textbf{a}}$~}
\newcommand{\baxis}{$\textit{\textbf{b}}$~}

\begin{document}

\title{Evolution of the nuclear and magnetic structures of TlFe$_{1.6}$Se$_2$ with temperature}

\author{Huibo Cao}
\affiliation{Quantum Condensed Matter Division, Oak Ridge National Laboratory, Oak Ridge, TN 37831}
\author{Claudia Cantoni}
\affiliation{Materials Science and Technology Division, Oak Ridge National Laboratory, Oak Ridge, TN 37831}
\author{Andrew F. May}
\affiliation{Materials Science and Technology Division, Oak Ridge National Laboratory, Oak Ridge, TN 37831}
\author{Michael A. McGuire}
\affiliation{Materials Science and Technology Division, Oak Ridge National Laboratory, Oak Ridge, TN 37831}
\author{Bryan C. Chakoumakos}
\affiliation{Quantum Condensed Matter Division, Oak Ridge National Laboratory, Oak Ridge, TN 37831}
\author{Stephen J. Pennycook}
\affiliation{Materials Science and Technology Division, Oak Ridge National Laboratory, Oak Ridge, TN 37831}
\author{Radu Custelcean}
\affiliation{Chemical Sciences Division, Oak Ridge National Laboratory, Oak Ridge, TN 37831}
\author{Athena S. Sefat}
\affiliation{Materials Science and Technology Division, Oak Ridge National Laboratory, Oak Ridge, TN 37831}
\author{Brian C. Sales}
\affiliation{Materials Science and Technology Division, Oak Ridge National Laboratory, Oak Ridge, TN 37831}

\begin{abstract}
The evolution of the nuclear and magnetic structures of TlFe$_{1.6}$Se$_2$ was determined in the temperature range of 5-450\,K using single crystal neutron diffraction.  The Fe layers in these materials develop a corrugation in the magnetically ordered state.  A canting away from the block checkerboard magnetic structure is observed in the narrow temperature range between approximately 100 and 150\,K.  In this same temperature range, an increase in the corrugation of the Fe layers is observed.  At lower temperatures, the block checkerboard magnetic structure is recovered with a suppressed magnetic moment and abrupt changes in the lattice parameters.  Microstructure analysis at 300\,K using atomic-resolution Z-contrast scanning transmission electron microscopy reveals regions with ordered and disordered Fe vacancies, and the iron content is found to be uniform across the crystal.  These findings highlight the differences between the alkali-metal and thallium materials, and indicate competition between magnetic ground states and a strong coupling of magnetism to the lattice in TlFe$_{1.6}$Se$_2$.
\end{abstract}


\maketitle

The recent discovery of superconductivity at temperatures greater than 30\,K in A$_x$Fe$_y$Se$_2$ is generating much effort to understand the relationship between superconductivity, magnetism, and structure in these materials.\cite{jgguo,Zavalij,afwang, krzton, mhfang, wbao1, wbao2, fye, park,yu,li,song,ricci,felser,yuan} These compounds crystallize in a variant of the ThCr$_2$Si$_2$-type structure, with vacancy ordering in the iron layer yielding a $\sqrt{5}a \times \sqrt{5}a$ supercell.  Although TlFe$_{1.6}$Se$_2$ is a magnetic insulator at low temperatures,\cite{sales11i} this compound is chemically and structurally similar to recently discovered superconducting compositions where part or all of the Tl is replaced by an alkali metal such as K, Cs, or Rb. \cite{wbao1, wbao2, fye, park}  For instance, similar to the superconducting compositions, vacancy ordering occurs below $T_v\approx~$460\,K in TlFe$_{1.6}$Se$_2$, and a block checkerboard antiferromagnetic magnetic order emerges below $T_N\approx~$430\,K.\cite{sales11i} There are significant differences, however, between the superconducting compositions and TlFe$_{1.6}$Se$_2$.

For TlFe$_{1.6}$Se$_2$ crystals grown from the melt, the Tl site is fully occupied, and the total Fe content shows little variation regardless of the starting composition.  It is likely, however, that the degree of vacancy order depends on the starting composition and the processing conditions. The superconducting compositions appear to be more flexible, and the best superconducting properties are obtained when the alkali metal and iron site occupancies are both near 80\%, e.g., K$_{0.8}$Fe$_{1.6}$Se$_2$.\cite{jgguo, afwang, krzton, mhfang, wbao1, wbao2, fye,sabrowsky,haggstrom} One distinct advantage of the TlFe$_{1.6}$Se$_2$ crystals is that they are stable in air for many hours and do not exhibit any tendency toward phase separation into Tl-rich and Tl-poor regions, which appears to occur when Tl is replaced by an alkali metal.\cite{feihan,yuan}

\begin{figure*}[!ht]
\includegraphics [width=7in]{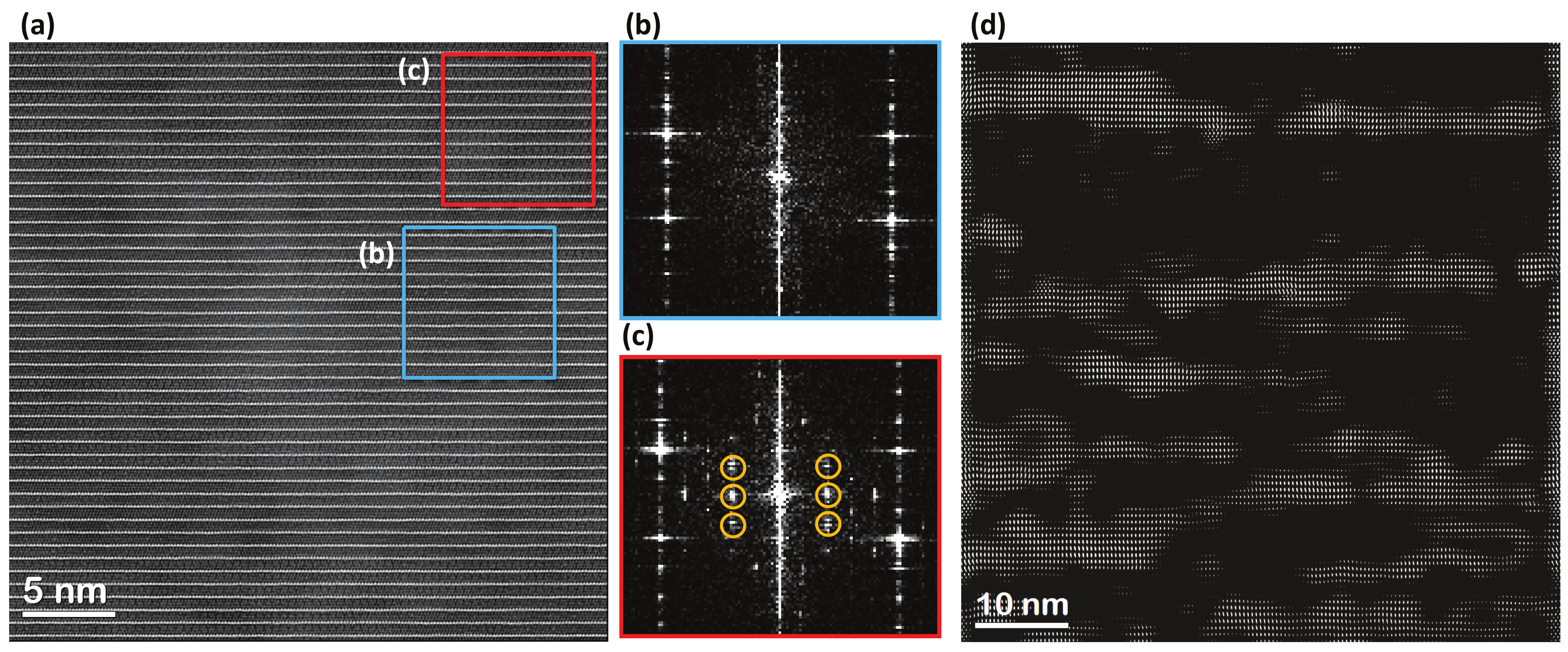}
\caption {(color online) (a) HAADF image with the electron beam parallel to the $\aaxis$-axis of the $\sqrt{5}a\times\sqrt{5}a$ supercell. The bright lines correspond to the heavier Tl planes. (b) and (c) are digital diffraction patterns from the regions outlined in (a) by blue (disordered Fe vacancies) and red (ordered Fe vacancies) squares, respectively. (d) Fourier filter obtained by selecting the superstructure reflections indicated by circles in (c), with black regions corresponding to disordered regions.}
\label{FFT}
\end{figure*}

Our previous work on TlFe$_{1.6}$Se$_2$ revealed two unexpected phase transitions, at $T_1\approx~$140\,K and $T_2\approx~$100\,K, which were evident in resistivity, magnetic susceptibility and heat capacity data.\cite{sales11i}  Upon cooling, the magnetic order parameter was observed to reach a maximum at $\sim T_1$, below which it decreased with a particularly sharp decrease just below $T_2$.  These transitions did not correspond to a change in the crystal symmetry or vacancy ordering, and understanding their physical origin was the motivation for this detailed structural study.

In the present work, the nuclear and magnetic structures in the vicinity of the two phase transitions $T_1$ and $T_2$ are examined in detail using single crystal neutron and X-ray diffraction.  The possibility of phase separation into fully-ordered and disordered regions is examined by atomic-resolution Z-contrast scanning transmission electron microscopy (Z-STEM).  Such phase separation has been reported in crystals and films of K$_x$Fe$_{2-y}$Se$_2$,\cite{feihan,wangz,li,song} and in Rb$_{0.8}$Fe$_{1.6}$Se$_2$.\cite{felser}

\begin{figure}[!ht]
\includegraphics [width=3.2in]{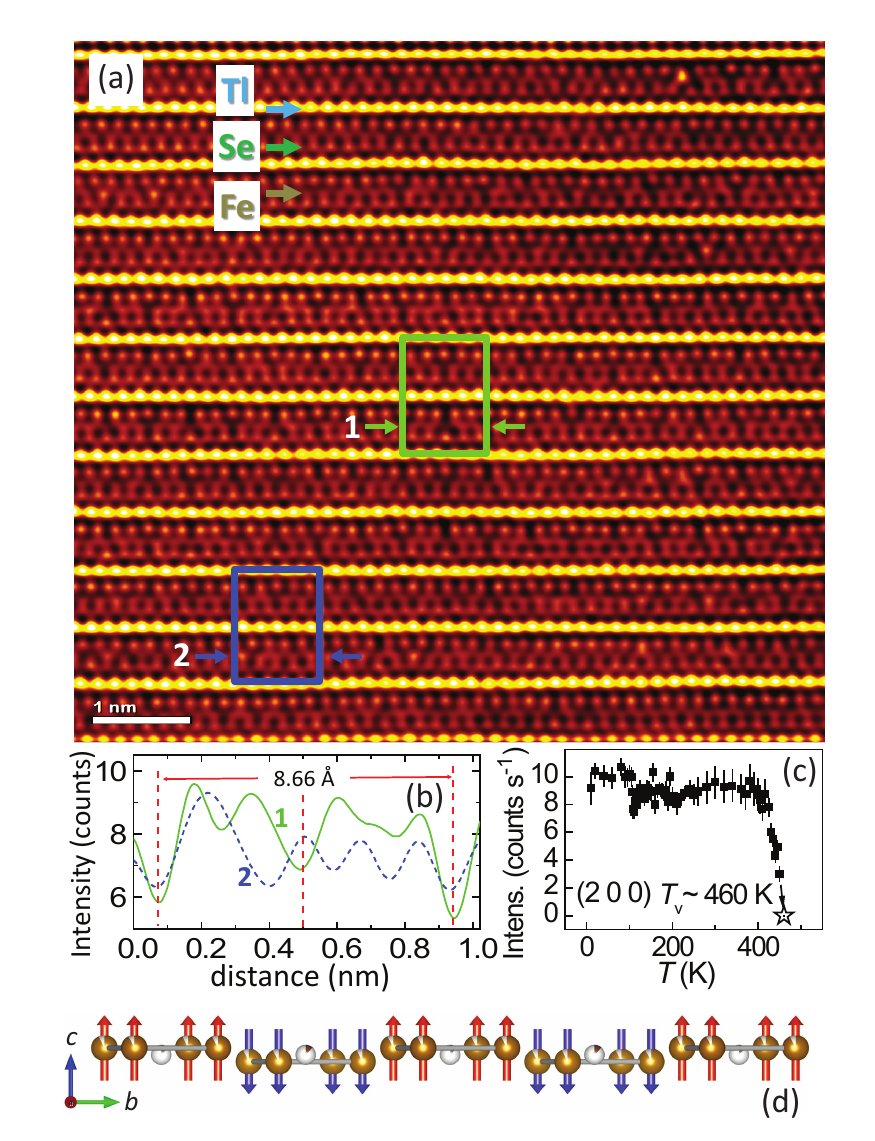}
\caption {(color online) (a) High magnification HAADF image with the electron beam parallel to a. The brightest planes are Tl planes. The nearest neighbor planes are Se planes, and the dimmest planes sandwiched between two Se planes are Fe planes. The green (blue) rectangle frames one unit cell. (b) Plot of the HAADF intensity along the Fe planes labeled 1 and 2 in (a). (c) Temperature dependent integrated intensity of the Bragg peak (2 0 0), which is related to vacancy order. (d) Projection of the Fe layer of TlFe$_{1.6}$Se$_2$ along the \baxis axis; spheres represent Fe where the Fe site occupancies are shown by partial brown coloring with Fe moments indicated by red and blue arrows to highlight the spin blocks.}
\label{tem}
\end{figure}

Single crystals of TlFe$_{1.6}$Se$_2$ were grown from a melt of nominal composition TlFe$_2$Se$_2$ using a modified Bridgman technique as discussed in Ref. \citenum{sales11i}.  The crystals were observed to be homogeneous by energy dispersive spectroscopy, with a resolution on the order of 50\,$\mu$m. Single crystal neutron diffraction measurements from 5\,K to 450\,K were made at HB-3A four-circle diffractometer at the High Flux Isotope Reactor at the Oak Ridge National Laboratory.  Complete neutron diffraction data sets were collected at 5\,K, 90\,K, 115\,K, 140\,K, 175\,K, 250\,K, 340\,K, and 450\,K.  The neutron wavelength of 1.536 \AA~ was used with a bent perfect Si-220 monochromator.\cite{chak} Single crystal X-ray diffraction was measured at 135\,K and 200\,K with a Bruker SMART APEX CCD diffractometer with Mo-K$\alpha$ radiation.  The neutron diffraction data refinements are based on $\sim$250 reflections and were completed in the program Fullprof.\cite{fullprof} The data were refined using the $\sqrt{5}a$$\times$$\sqrt{5}$a superlattice notation, and both nuclear and magnetic structures were resolved.  Refinement results from neutron data at selected temperatures are shown in Table \ref{tab:refine}.  See Ref. \citenum{sales11i} for additional information.

The microstructure of the crystals was examined at 300\,K using atomic-resolution Z-contrast scanning transmission electron microscopy (Z-STEM).  Aberration-corrected high angle annular dark field (HAADF) imaging was performed in a Nion UltraSTEM equipped with cold field emission gun, third generation C3 and C5 aberration corrector, and operated at 100\,kV.  Samples were thinned to nearly 20 nm via dual focused ion beam (FIB) slicing followed by Ar ion milling at a voltage of 0.5\,kV. FIB was employed as an alternative to conventional grinding and polishing to avoid introducing excessive strain in the sample. Low voltage ion milling was employed to remove FIB induced amorphization of the sample surface.  The image was treated using a probe deconvolution algorithm to improve contrast and reduce noise.\cite{tem}

Previous refinement of the neutron diffraction data revealed the magnetic moment to be 1.7\,$\mu_B$/Fe at room temperature in TlFe$_{1.6}$Se$_2$.\cite{sales11i} This is smaller than observed in the analogous superconducting compositions, where the moment is $\gtrsim$3\,$\mu_B$/Fe below $\sim$250\,K.\cite{fye}  Additionally, multiple, unexplained low-temperature magnetic phase transitions were observed in the magnetic susceptibility, electrical resistivity and heat capacity of TlFe$_{1.6}$Se$_2$ at $T_1$ and $T_2$.\cite{sales11i}  Electron microscopy was employed to determine if these results could be explained by phase separation into regions of differing vacancy order, which would likely have different magnetic behaviors.\cite{felser}  

\begin{table}
\caption{Refined structural parameters for TlFe$_{1.6}$Se$_2$ from single-crystal neutron diffraction data at three temperatures; space group $I4/m$ (No. 87).  All non-listed atomic coordinates are zero except for Fe2, which is located at ($0,\frac{1}{2},z$); data were refined with $U_{eq}^{Tl1}$=$U_{eq}^{Tl2}$ and $U_{eq}^{Se1}=U_{eq}^{Se2}$.}
\begin{tabular}[c]{ccccccc}
\hline							
	&	5\,K	&	115\,K	&	250\,K	\\
\hline			&		&		\\
$a (\AA)$	&	8.641(4)	&	8.665(4)	&	8.685 (4)	\\
$c (\AA)$	&	13.993(7)	&	13.957(7)	&	14.002(7)	\\
\hline							
$x_{Tl2}$	&	0.1954(13)	&	0.1964(15)  	&	0.1943(18)  	\\
$y_{Tl2}$	&	0.3957(11)	&	0.3959(15)  	&	0.3966(14)  	\\
$x_{Fe1}$	&	0.0928(7)	&	0.0933(7)   	&	0.0926(8)   	\\
$y_{Fe1}$	&	0.1970(7)	&	0.1964(8)   	&	0.1967(8)   	\\
$z_{Fe1}$	&	0.2476(6)	&	0.2452(9)   	&	0.2485(8)   	\\
$z_{Fe2}$	&	0.727(5)  & 0.717(5)      & 0.722(5)   	\\
$z_{Se1}$	&	0.3651(12)	&	0.3638(14)  	&	0.3648(15)  	\\
$x_{Se2}$	&	0.1986(10)	&	0.1980(11)  	&	0.1993(12)  	\\
$y_{Se2}$	&	0.3923(8)	&	0.3925(10)  	&	0.3921(9)   	\\
$z_{Se2}$	&	0.3565(3)	&	0.3560(4)   	&	0.3557(4)   	\\
\hline					
$U_{eq}^{Tl1}\, (\AA^2)$	&	0.0184(16)	&	0.0295(17)  	&	0.038(2)    	\\
$U_{eq}^{Fe1}\, (\AA^2)$	&	0.0126(18)	&	0.0164(20)  	&	0.020(2)    	\\
$U_{eq}^{Fe2}\, (\AA^2)$	&	0.04(2)	&	0.07(3)     	&	0.05(3)     	\\
$U_{eq}^{Se1}\, (\AA^2)$	&	0.0185(14)	&	0.0244(15)  	&	0.0261(18)  	\\
\hline							
$R_1$	&	0.0538	&	0.0522	&	0.0589	\\
$wRF^2$	&	0.122	&	0.123	&	0.132	\\
$\chi^2$	&	7.7	&	9.11	&	8.04	\\
\hline							
\end{tabular}
\label{tab:refine}
\end{table}

Examination of the microstructure in TlFe$_{1.6}$Se$_2$ using cross section Z-STEM reveals regions of vacancy order and disorder, as observed in Figure \ref{FFT}. Figure \ref{FFT}a is an HAADF image with the electron beam directed along the $\aaxis$ axis of the $\sqrt{5}a\times \sqrt{5}a$ supercell. A close inspection of Fig. \ref{FFT}a shows the existence of two regions yielding slightly different contrast and lattice periodicity.  While regions like the one highlighted by the blue square labeled (b) yield fast Fourier transforms (FFT) that are consistent with a ThCr$_2$Si$_2$-type structure (Fig. \ref{FFT}b), regions like the one framed by the red square show superlattice reflections associated with periodic minima of intensity in the Fe plane (see Fig. \ref{FFT}c).  This allows the ordered and disordered regions to be visualized, as in Fig. \ref{FFT}d, which is a filtered image obtained by selecting only the frequencies yielding superstructure reflections indicated by the circles in Fig. \ref{FFT}c. Figure \ref{FFT}d highlights the spatial extent of those regions producing superstructure reflections, clearly delineating the ordered (light) regions from the disordered (dark) regions. Using Fig. \ref{FFT}d, and similar images (spanning areas of 64 $\times$ 64\,nm$^2$), a volume fraction in the range 40 –- 48\% is obtained for the ordered regions.  This quantification assumes that the ordered (disordered) regions extend for the entire thickness of the TEM specimen (nearly 20\,nm) in the direction perpendicular to the image plane.

Figure \ref{tem}a is a higher magnification HAADF image of the same sample. The center and right hand corners of the image have highly ordered vacancies, while the left hand corners are disordered. The green and blue rectangles highlight unit cells in the ordered and disordered regions, respectively.

Intensity line profiles along the Fe planes labeled 1 and 2 in Fig. \ref{tem}a are reported in Fig. \ref{tem}b.  These reveal the location of vacant Fe sites as positions of the deep minima in intensity. Curve 1 exhibits deep minima in correspondence with the highly vacant Fe2 sites of the model in Fig. \ref{tem}d, as expected for a well ordered region with the $\sqrt{5}a\times\sqrt{5}a$ supercell. Curve 2 is not entirely consistent with the model, as expected for a disordered distribution of vacancies. Analysis of intensity line profiles for the Fe planes over distances of $\sim$35\,nm suggests that the ordered and disordered regions have on average the same number of vacancies.  Therefore, the ordered and disordered regions have the same composition and differ only by the arrangement of the Fe atoms (vacancies).


In summary, the Z-STEM data suggest a distribution of regions of $\sim$10\,nm--30\,nm in size along the basal plane and 2--4 unit cells along the \caxis-axis having a high concentration of ordered Fe vacancies.  The Z-STEM also reveals these ordered/disordered regions are crystallographically coherent, which is consistent with the observation of sharp diffraction peaks.

The presence of ordered and disordered regions does not entirely explain the observed physical properties.  Specifically, the observation of two magnetic transitions cannot be understood as one transition occurring in the ordered region and the other transition occurring in the disordered region. Such a conclusion is inconsistent with the diffraction results discussed below.   However, an interaction between these regions, most likely via lattice strain, may be responsible for the observed behavior.

Single crystal neutron and X-ray diffraction both reveal a vacancy ordering transition around 460\,K.  The temperature dependent integrated neutron intensity of the nuclear Bragg peak (2 0 0) is plotted in Fig.\ref{tem}c. The (2 0 0) reflection is solely due to ordering of the Fe vacancies and fully disappears at $\sim$ 460\,K by extrapolating the observed data points. Upon cooling to $\sim$ 400\,K, the vacancy order saturates.  Therefore, the Z-STEM data at 300\,K are expected to be representative of the microstructures at lower temperatures as well.  

All of the diffraction data are well described by a $\sqrt{5}$\aaxis $\times$ $\sqrt{5}$\aaxis superstructure with a highly Fe-occupied position Fe1 (on average 93\% occupied) and a highly-vacant position Fe2 (on average 27\% occupied), consistent with our prior report.\cite{sales11i} The volume percentage of ordered and disordered regions estimated from the HAADF images is consistent with these average occupancies.  The refinement results are summarized in Table \ref{tab:refine} for $T$ = 5\,K, 115\,K, and 250\,K.  

The vacancy ordered Fe layer is shown in Fig.\ref{tem}d, with Fe occupancies proportional to the filling of the brown spheres.  The Fe moments in the block checkerboard structure are shown as red and blue arrows to highlight the blocks of spins, which are ferromagnetically aligned within a block and antiferromagnetic between blocks.   The corrugation ($\delta$) between the spin blocks in Fig. \ref{tem}d is temperature dependent and will be discussed in detail below; $\delta$ is defined as the distance along $\caxis$ that separates the spin blocks.

\begin{figure}[!ht]
\includegraphics* [width=3.2in]{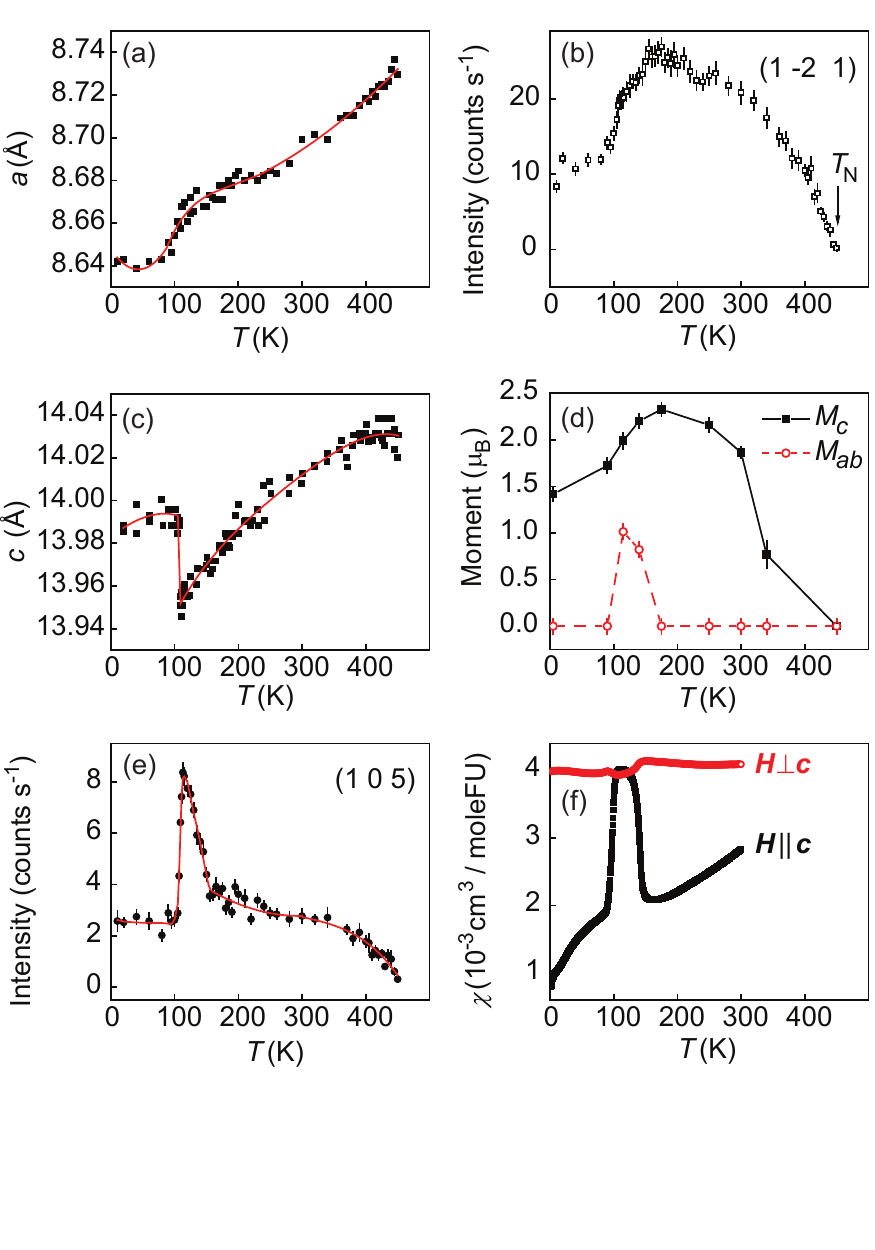} 
\caption {In (a) and (c) the lattice parameters are shown to respond strongly to the magnetic transitions, particularly near 100\,K.  (b) A magnetic order parameter (intensity of (1 -2 1) peak) that is influenced by moments aligned with \caxis reveals a suppressed moment at low $T$. (d) Projection of the magnetic moment per Fe onto the \textit{ab}-plane and along \caxis demonstrates the canting of the moment at 115 and 135\,K. (e) The temperature dependence of the (1 0 5) peak reveals a structural anomaly in the temperature region 100-140\,K, which is primarily associated with corrugation of the Fe layers. (f) Anisotropic magnetic susceptibility\cite{sales11i} for single crystalline TlFe$_{1.6}$Se$_2$ are consistent with a canting of the moments away from \caxis between $T_1\approx$100 and $T_2\approx$140\,K.}
\label{6panel}
\end{figure}

Temperature dependent diffraction was utilized to study the origin of phase transitions at $T_1\,\approx$\,140\,K and $T_2\,\approx$\,100\,K, which were observed by measurements of the magnetic susceptibility (reproduced in Fig. \ref{6panel}f), electrical resistivity, and heat capacity.\cite{sales11i} Previously, we determined the \caxis lattice parameter variation for a single-crystal using a 2-circle X-ray diffractometer, which showed behavior similar to that presented in Fig. \ref{6panel}c.\cite{sales11i}  Here, we show the whole temperature range for both lattice parameters determined with a four-circle neutron diffractometer, and the refinement results show a clear structural transition at $\sim$100\,K.  As observed in Fig. \ref{6panel}a,c, \caxis increases abruptly near 100\,K, while \aaxis decreases rapidly near 100\,K and has unusual temperature dependence between 100 and 200\,K.  No similar sharp anomaly was observed in unit cell volume at 100\,K.  The anomalous temperature dependence of \aaxis and \caxis above $\sim$200\,K is likely due to the nearby order-disorder transition at 460\,K.

\begin{figure}[!ht]
\includegraphics* [width=3.2in]{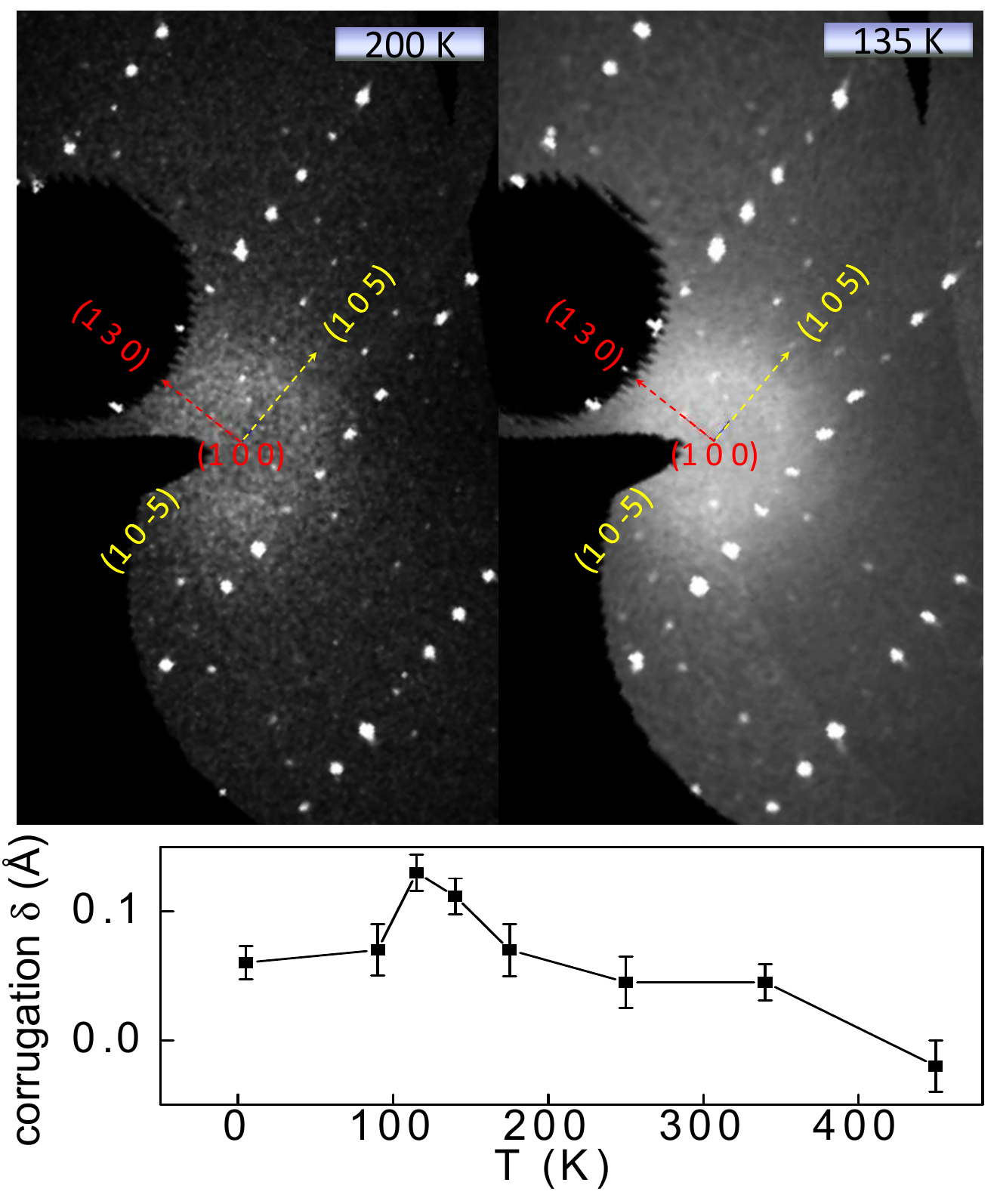}  
\caption {(color online) Single crystal X-ray diffraction patterns with [1 0 0] cut showing the (1 0 5) and (1 0 -5) Bragg peaks have much larger intensity at 135\,K compared to 200\,K.  This is associated with an increased corrugation ($\delta$) of the Fe layer, which is obtained from refinements of the neutron diffraction data.  $\delta$ is defined as the distance along \caxis between the spin blocks shown in Fig. \ref{tem}d.}
\label{xray}
\end{figure}

In the magnetically ordered state, the Fe layer is no longer flat but develops a slight corrugation with the block of 4 up spins (down spins) moving up (down).  The parameter $\delta$ is used to quantify this corrugation, and is given by the distance along \caxis that separates the Fe spin blocks.  In the absence of corrugation, Fe1 would be found at $z$=$\frac{1}{4}$.  Therefore, $\delta/c$ is equal to twice the deviation from this ideal position of Fe1.  At 340 and 250\,K, refinements yield $\delta\,\approx\,0.04\,$\AA~.  The temperature dependence of the corrugation obtained by refining full diffraction data sets is shown in Fig. \ref{xray}.  The corrugation is only present in the magnetic state, as observed by a non-zero value below $T_N$, which indicates a strong coupling of magnetism to the lattice.  The corrugation begins to increase rapidly near $T_1$, and reaches a maximum value of $\delta\,\approx\,0.13\,$\AA~ at 115\,K before a rapid decrease at $T_2$ to a value of $\delta\,\approx\,0.07\,$\AA~, which is similar to that observed near room temperature.  As observed in Table \ref{tab:refine}, the structural changes across $T_1$ and $T_2$ are dominated by changes in the lattice constants and changes in the corrugation.  The position of Fe1 in Table \ref{tab:refine} clearly demonstrates the corrugation, and the corresponding influence on Se1 (located above or below a spin block) is also be observed.

The temperature dependence of the corrugation can be tracked via the intensity of the (1 0 5) peak, which partly reflects the corrugation and vanishes in the limit of zero corrugation.  The (1 0 5) peak also gains intensity through a canting of the moment, which is discussed below. X-ray diffraction is therefore necessary to examine the extent to which the temperature dependence of the (1 0 5) peak is related to changes in the nuclear structure.  Single crystal X-ray diffraction was performed at 200\,K and 135\,K, and reciprocal space maps with [1 0 0] cut are shown in Fig. \ref{xray}.  In these reciprocal space maps, one can clearly see the Bragg peak (1 0 5) and its equivalent (1 0 -5) at 135\,K but not at 200\,K.  To quantify the increase in intensity of (1 0 5) related to changes in the nuclear structure between 200 and 135\,K, the intensity of (1 0 5) is integrated and normalized to the integrated intensity of (2 0 0), which is independent of temperature over this range (see Fig. \ref{tem}c).  The relative integrated intensities int(1 0 5)/int(2 0 0) are 0.43(8) at 200\,K and 1.17(8) at 135\,K.  This proves the nuclear contribution at (1 0 5) rises significantly between 200\,K and 135\,K, and thus the magnetic contribution to the data presented in Fig. \ref{6panel}e is relatively small.

The phase transitions at $T_1$ and $T_2$ have clear structural components.  The transition at 100\,K appears particularly sharp, as evidenced by the jump in \caxis (Fig. \ref{6panel}c) and the abrupt decrease in the corrugation (Fig. \ref{xray}), while the structural effects that begin at $\sim$150\,K occur more gradually (Fig. \ref{6panel}a,c).  Both phase transition temperatures are well defined in the magnetic susceptibility data (Fig. \ref{6panel}f), and the behavior of the magnetic structure through $T_1$ and $T_2$ are now considered.\cite{sales11i}

The (1 -2 1) peak shown in Fig. \ref{6panel}b is utilized as a magnetic order parameter.  This reflection is purely magnetic, as indicated by its vanishing intensity at $T_N$, and is sensitive to the \caxis component of the moment.  The intensity of (1 -2 1) tracks that of the moment obtained by complete refinement, which is shown in Fig. \ref{6panel}d.  For the majority of temperatures, the total moment is equal to the moment along the \caxis-axis ($M_c$).  Upon cooling, this moment reaches a maximum at $\sim$150\,K and then begins to decrease.  A clear feature is observed near $T_2$, below which the intensity of (1 -2 1) decreases rapidly.

The magnetism is naturally broken into three temperature regions.\cite{sales11i} Upon cooling below $T_N\sim$430\,K, a block-checkerboard antiferromagnetic structure sets in with spins aligned along \caxis.  As expected for an antiferromagnetic system, the magnetic order parameter increases upon cooling below $T_N$ (Fig. \ref{6panel}b).  Similarly, the magnetic susceptibility ($\chi$ in Fig. \ref{6panel}f) is relatively independent of $T$ for fields (\textbf{\textit{H}}) perpendicular to the spin orientation, while $\chi$ decreases below $T_N$ for \textbf{\textit{H}}$\parallel$\caxis.  This behavior is maintained down to about 150\,K, where the order parameter reaches a maximum and begins to decrease. Interestingly, this corresponds to an increase in $\chi$ for \textbf{\textit{H}}$\parallel$\caxis and a decrease in $\chi$ for \textbf{\textit{H}}$\perp$\caxis. Upon cooling below $\sim$100\,K, the rate of decrease in the order parameter increases, and $\chi$ again behaves in a manner consistent with the observations above $\sim$150\,K.  The data for $\chi$ are taken from Ref. \citenum{sales11i} and were collected using a field of 5\,T (zero-field cooled).

\begin{figure}[!ht]
\includegraphics* [width=3.2in] {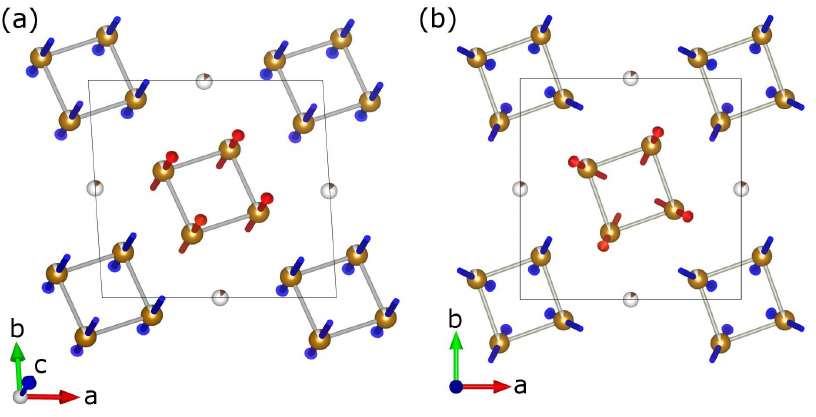} 
\caption {(color online)(a) Block-checkerboard antiferromagnetic order with collinear Fe moments along \caxis and (b) canted Fe moments as determined by neutron diffraction at 115\,K.}
\label{spins}
\end{figure}

At the temperatures above 150\,K and below 100\,K, the block-checkerboard antiferromagnetic model with the moments along \caxis provides a good description of the neutron diffraction data. At temperatures between 100\,K and 150\,K, however, in-plane components of the moment are required.  A canting of the moment is consistent with the trends in magnetic susceptibility shown in Fig. \ref{6panel}f.  The resulting block checkerboard and canted spin structures are shown in Fig. \ref{spins}, where only the Fe atoms and moments are shown for clarity.

Through refinements of complete neutron data sets, the total moment is found to saturate to 2.3(1)\,\mub at 150\,K and maintains this value until roughly 115\,K, while the \caxis-axis component of the Fe moment drops to 2.0(1)\,\mub.  Between 100\,K and 150\,K, an \textit{ab}-plane component of the moment ($M_{ab}$) is observed.  As shown in Fig. \ref{6panel}d, $M_{ab}$ is only non-zero in the region between 100\,K and 150\,K, which corresponds with the changes in $\chi$ and the corrugation. The angle between the moment and the \caxis-axis is 20(2)$^{\circ}$ at 140\,K and 27(3)$^{\circ}$ at 115\,K. Upon further cooling to 100\,K, the block-checkerboard structure is recovered, but with a suppressed moment of 1.4(1)\,$\mu_B$, as shown in Fig. \ref{6panel}d.  Given the saturation of the vacancy ordering at high temperatures, this temperature-dependence of the moment is unlikely to be caused by changes in the degree of vacancy ordering.

The magnetic ordering may be restricted to the vacancy ordered regions, as proposed for the alkali metal compounds.\cite{felser,yuan}  However, the measurements presented here do not allow the magnetic behavior of the vacancy ordered and disordered regions to be determined independently.

Based on the results presented above, the temperature dependence of the crystal and magnetic structure of TlFe$_{1.6}$Se$_2$ can be summarized as follows.  From 430--150\,K the magnetic moments are ordered in the block-checkerboard arrangement with moments lying along the \caxis direction. The Fe layer shows a slight corrugation in this temperature range. Upon cooling, at $T_1\approx140$\,K the moments begin to cant away from the \caxis-axis. As the moments cant, the corrugation of the Fe layer is significantly increased, indicating strong magnetoelastic interactions. This continues down to $T_2\approx100$\,K, at which point the canting abruptly vanishes and the higher temperature magnetic structure is recovered. However, the magnitude of the moment below $T_2$ is significantly reduced from the value above $T_1$. These changes in the magnetism upon cooling through $T_2$ are accompanied by two notable crystallographic changes: the corrugation sharply decreases and the \caxis-lattice constant sharply increases, both reaching values similar to those seen above $T_1$.

The observed increase in \caxis and decrease in the magnetic moment from a maximum of 2.3(1)\,\mub near 150\,K to 1.4(1)\,\mub below $T_2$ are remarkable. It is particularly unusual to find a smaller magnetic moment at lower temperature with no difference in magnetic structure. Orbital ordering is predicted to occur in Fe layers of TlFe$_{1.6}$Se$_2$ when the vacancies are fully ordered,\cite{wlv} and provides a simple explanation for the observed behavior. In the absence of orbital ordering, each Fe has about two $d$-electrons distributed among three nearly degenerate $t_{2g}$ levels (with two lower lying, filled, $e_{g}$ levels) giving a moment of about 2\,$\mu_B$ per Fe. In the orbitally-ordered state proposed by Lv et al.,\cite{wlv} the $t_{2g}$ orbitals are split so that each Fe has either $d_{xz}$ or $d_{yz}$ as the lowest energy level. Thus, orbital-ordering would decrease the occupancy of the $d_{xy}$ orbital as the electrons fill the lower energy states. The resulting increase in population of the out of plane ($d_{xz}$ and $d_{yz}$) orbitals is consistent with the observed increase in \caxis due to electrostatic repulsion. In addition, double occupation of the lowest lying $t_{2g}$ orbital would result in a decrease in the magnetic moment. With two $t_{2g}$ electrons per iron and complete orbital-ordering, the moment would be expected to vanish. However, the observation of a non-zero moment at low temperatures is not surprising, since incomplete vacancy ordering would likely preclude perfect orbital ordering.  

This structural investigation of TlFe$_{1.6}$Se$_{2}$ reveals that two low temperature phase transitions originate from competition between magnetic ground states and the coupling of magnetism to the lattice. The observation of competing ground states in TlFe$_{1.6}$Se$_2$ highlights the complexity of the intercalated iron-selenide compounds.

Research at Oak Ridge National Laboratory's High Flux Isotope Reactor was sponsored by the Scientific User Facilities Division, Office of Basic Energy Sciences, U. S. Department of Energy. The research was also supported in part by the Materials Sciences and Engineering Division (BCS, MAM, AFM, CC, SJP, and AS), Office of Basic Energy Sciences, U.S. Department of Energy.  RC was supported by the Division of Chemical Sciences, Geosciences, and Biosciences, Office of Basic Energy Sciences, U.S. Department of Energy.


\begin{thebibliography}{}

\bibitem{jgguo} J. G. Guo, S. F. Jin, G. Wang, S. C. Wang, K. X. Zhu, T. T. Zhou, M. He, and X. L. Chen, Phys. Rev. B {\bf 82}, 180520(R) (2010).
\bibitem{Zavalij}P. Zavalij, W. Bao, X. F. Wang, J. J. Ying, X. H. Chen, D. M. Wang, J. B. He, X. Q. Wang, G. F. Chen, P.-Y. Hsieh, Q. Huang, and M. A. Green, Phys. Rev. B {\bf 83}, 132509 (2011).
\bibitem{afwang} A. F. Wang, J. J. Ying, Y. J. Yan, R. H. Liu, X. G. Luo, Z. Y. Li, X. F. Wang, M. Zhang, G. J. Ye, P. Cheng, Z. J. Xiang, X. H. Chen, Phys. Rev. B {\bf 83}, 060512 (2011).
\bibitem{krzton} A. Krzton-Maziopa, Z. Shermadini, E. Pomjakushina, V. Pomjakushin, M. Bendele, A. Amato, R. Khasanov, H. Luetkens, K. Conder,  J. Phys.: Condens. Matter {\bf 23}, 052203 (2011). 
\bibitem{mhfang} M. H. Fang, H. D. Wang, C. H. Dong, Z. J. Li, C. M. Feng, J. Chen, H. Q. Yuan, Europhys. Lett. {\bf 94}, 27009 (2011).
\bibitem{wbao1} W. Bao, Q. Huang, G. F. Chen, M. A. Green, D. M. Wang, J. B. He, X. Q. Wang, Y. Qiu, Chinese Phys. Lett. {\bf 28}, 086104 (2011).
\bibitem{wbao2} Wei Bao, G. N. Li, Q. Huang, G. F. Chen, J. B. He, M. A. Green, Y. Qiu, D. M. Wang, J. L. Luo, arXiv: 1102.3674 (2011).
\bibitem{fye} F. Ye, S. Chi, Wei Bao, X. F. Wang, J. J. Ying, X. H. Chen, H. D. Wang, C. H. Dong, Minghu Fang, Phys. Rev. Lett. {\bf 107}, 137003 (2011).
\bibitem{park}	J. T. Park, G. Friemel, Yuan Li, J.-H. Kim, V. Tsurkan, J. Deisenhofer, H.-A. Krug von Nidda, A. Loidl, A. Ivanov, B. Keimer, D. S. Inosov, arXiv:1107.1703v1 (2011).
\bibitem{yu} R. Yu, P. Goswami, and Q. Si, Phys. Rev. B. {\bf 84}, 094451 (2011).
\bibitem{li} W. Li, H. Ding, P. Deng, K. Chang, C. Song, K. He, L. Wang, X. Ma, J-P Hu, X. Chen, and Q-K Xue, Nature Physics, \textit{advanced online publication; doi:10.1038/nphys2155} (2011).
\bibitem{song} Y. J. Song, Z. Wang, Z.W. Wang, H. L. Shi, Z. Chen, H. F. Tian, G. F. Chen, H. X. Yang and J. Q. Li, Europhys. Lett. {\bf 95}, 37007 (2011)
\bibitem{ricci} A. Ricci, N. Poccia, G. Campi, B. Joseph, G. Arrighetti, L. Barba, M. Reynolds, M. Burghammer, H. Takeya, Y. Mizuguchi, Y. Takano, M. Colapietro, N. L. Saini, and A. Bianconi, Phys. Rev. B {\bf 84}, 060511(R) (2011)
\bibitem{felser} V. Ksenofontov, G. Wortmann, S. A. Medvedev, V. Tsurkan, J. Deisenhofer, A. Loidl, and C. Felser,  Phys. Rev. B {\bf 84}, 180508(R) (2011)
\bibitem{yuan} R. H. Yuan, T. Dong, Y. J. Song, P. Zheng, G. F. Chen, J. P. Hu, J.Q. Li and N. L. Wang, Sci. Rep. {\bf 2} doi:10.1038/srep00221 (2012)
\bibitem{sales11i}	Brian C. Sales, Michael A. McGuire, Andrew F. May, Huibo Cao, Bryan C. Chakoumakos, and Athena S. Sefat, Phys. Rev. B {\bf 83}, 224510 (2011).
\bibitem{fullprof} J Rodríguez-Carvajal, Physica B {\bf 192} 55 (1993).
\bibitem{sabrowsky}	H. Sabrowsky, M. Rosenberg, D. Welz, P. Deppe, and W. Sch$\rm \ddot{a}$fer, J. Magn. Magn. Mat. {\bf 54}, 1497 (1986).
\bibitem{haggstrom} L. H$\rm \ddot{a}$ggstr$\rm \ddot{o}$m, H. R. Verma, S. Bjarman, R. W$\rm \ddot{a}$ppling, and R. Berger, J. Solid State Chem. {\bf 63} 401 (1986).
\bibitem{wangz} Z. Wang, H. L. Shi, Z. W. Wang, Z. Chen, H. F. Tian, G. F. Chen, J. G. Guo, H. X. Yang, and J. Q. Li, Phys. Rev. B {\bf 83}, 140505(R) (2011).
\bibitem{feihan}	F. Han, B. Shen, Z. Y. Wang, H. H. Wen, arXiv:1103.1347 (2011).
\bibitem{chak} B. C. Chakoumakos, H. Cao, F. Ye, A. D. Stoica, M. Popovici, M. Sundaram, W. Zhou, J. S. Hicks, G. W. Lynn, and R. A. Riedel, J. Applied Cryst., {\bf 44}, 655 (2011).
\bibitem{tem} http://tem-s3.nano.cnr.it/software.
\bibitem{wlv} Weicheng Lv, Wei-Cheng Lee, and Philip W. Phillips, Phys. Rev. B {\bf 84}, 155107 (2011).

\end{thebibliography}
\end{document}